\documentclass[amsmath,amssymb,prc,final,showpacs,twocolumn]{revtex4-1}
 
\usepackage{bm,hyphenat,xspace}
\usepackage{graphicx,epsfig}

\newcommand {\anz}{a_n^-}
\newcommand {\ankz}{a_{n,k}^-}

\newcommand {\mcu}{\mathcal{U}}

\begin{document}

\title {Efimov resonances above four-boson threshold}
  
\author{A.~Deltuva} 
\affiliation
{Institute of Theoretical Physics and Astronomy, 
Vilnius University, Saul\.etekio al. 3, LT-10257 Vilnius, Lithuania
}

\received{March 7, 2021} 

\begin{abstract}
Four-boson Efimov physics is well known in the negative energy regime but far less
above the four-body breakup threshold. The part of this region with negative two-boson scattering length
is studied solving rigorous four-particle scattering equations for transition operators
in the momentum space. Moving away from the unitary limit the Efimov tetramers evolve
from unstable bound states into resonances. Their energies and widths 
 are studied as functions of  the two-boson scattering length; 
a  universal behavior is established and  given in a dimensionless representation.
The Efimov tetramers have finite width in the whole regime; they broaden rapidly in the resonance regime
but remain narrower than the associated trimer.
 The resonant behavior is most clearly seen in the four-particle recombination rate.
\end{abstract}

 \maketitle

\section{Introduction \label{sec:intro}}

Efimov physics in few-body systems manifests itself by a rich spectrum of bound and/or resonant
states independent of the short-range interaction details.
 First, considering the three-body system with large
two-body scattering length $a$,  V. Efimov \cite{efimov:plb} predicted an infinite number of bound three-body states,
called Efimov trimers, with geometric energy spectrum in the unitary limit $a \to \infty$.  
After this  pioneering study  a large number of theoretical
and experimental works emerged as summarized in several review papers
\cite{braaten:rev,hammer:10a,naidon:rev,greene:rev,kievsky:rev21},
covering applications from nuclear to molecular and cold atom physics. 

Beside the most extensively investigated three-body systems,  also  four-body systems
attracted considerable interest. In particular, in the system of four identical bosons with 
resonant $s$-wave interactions it was found that for each Efimov trimer  there 
exist two four-boson states \cite{hammer:07a}, often called Efimov tetramers.  Except for the lowest two,
all the other are unstable bound states since they are above the ground state trimer. In the negative energy regime
below the
four-body breakup threshold the properties of the Efimov tetramers, i.e., the evolution of their energies
and widths with the scattering length, have been accurately calculated. At particular positive values of the 
scattering length $a$ the tetramers are crossing the particle-trimer or dimer-dimer  threshold and becoming
inelastic virtual states \cite{deltuva:12d}. On the negative $a$ side the tetramers decay through the four-particle
threshold of vanishing energy $E=0$. Each intersection of the tetramer and $E=0$ threshold taking place at specific
value of $a$ leads to a resonant enhancement of the four-particle recombination
\cite{stecher:09a,deltuva:12a}.

Once the tetramers enter the positive energy regime, they become true resonances, however,
their energies and widths quantitatively are not known.
Their study was precluded by the complications in rigorously describing the four-particle continuum 
with open many-cluster channels. Typically, in the coordinate-space differential equation approach
complicated boundary conditions must be imposed \cite{lazauskas:15a},
while in the momentum-space integral equation framework
one faces complicated singularities \cite{deltuva:12c}.
On top of that, the Efimovian character means that
the problem involves the states of very different sizes and energies,
that require a very careful and accurate treatment.

Recent developments in the description of four-nucleon reactions above the breakup threshold \cite{deltuva:12c},
especially the search of the four-neutron resonance \cite{deltuva:18b}, paved the way also for the study of 
four-boson Efimov resonances. Momentum-space integral equations for transition operators
with complex-energy and special weight method for the singularity treatment  is expected to provide
an accurate description of the four-boson system at positive energies, and will be used in the present work,
aiming to determine the properties of four-boson Efimov resonances and their impact on collision processes. 
The present study can also be viewed as an extension of the previous work \cite{deltuva:20a}
that considered the Efimov resonances in the three-boson system.

Section II shortly recalls the four-particle scattering equations
and essential aspects of calculations whereas results are given in Sec. III.
 The summary is presented in Sec. IV.

\section{Theory \label{sec:eq}}

Resonance corresponds to the pole of the $S$-matrix or the related transition operator
in the unphysical sheet of the complex-energy plane. Its location  $E^R - i\Gamma/2$ 
is determined by the real part of the energy $E^R$ and the width $\Gamma$. 
The Laurent series expansion for the dependence 
of the transition operator on the energy $E$ is led by the pole term proportional
to $1/(E-E^R + i\Gamma/2)$,  but higher-order  background terms  contribute as well.
As long as $\Gamma$ is not too large, the pole is not too far from the real
energy axis and therefore affects physical processes in the system at 
energies around $E^R$. Under these conditions the resonance parameters can be determined from
the energy dependence of the transition operator in the physical region,
i.e., from the physical observables. This procedure was used for the search of the
four-neutron resonance \cite{deltuva:18b} and Efimov three-boson resonances 
\cite{deltuva:20a}; it is described in more details in 
Refs.~\cite{deltuva:18b,deltuva:20a} and is used also in the present work.

The study of the four-neutron resonance \cite{deltuva:18b} considered 
$4 \to 4$ transition operator. It could be used also here, however, a more
practical choice is a subset of two-cluster transition operators, that are
more directly related to amplitudes for boson-trimer elastic and inelastic scattering
or four-boson recombination. These transition operators $\mcu_{\alpha \beta}(Z)$
are obtained solving
 symmetrized form of equations originally proposed by
Alt, Grassberger, and Sandhas (AGS) \cite{grassberger:67}, i.e., 
\begin{subequations} \label{eq:U}
\begin{align}  \nonumber
\mcu_{11}(Z)  = {}&  P_{34} (G_0  t  G_0)^{-1}  
+ P_{34}  U_1 G_0  t G_0  \mcu_{11}(Z)  \\
{}& + U_2 G_0  t G_0  \mcu_{21}(Z) , 
\label{eq:U11} \\  
\mcu_{21}(Z)  = {}&  (1 + P_{34}) (G_0  t  G_0)^{-1}  
+ (1 + P_{34}) U_1 G_0  t  G_0  \mcu_{11}(Z) , \label{eq:U21}
\end{align}
\end{subequations}
where the subscripts $\alpha,\beta= 1$ (2) denote the 3+1 (2+2) clustering,
$G_0 = (Z-H_0)^{-1}$ is the free four-boson resolvent of the
four-particle system with kinetic energy operator $H_0$ and $Z=E+i0$,
$t= v + v G_0 t $ is the two-boson transition matrix
derived from the pair (12)  potential $v$,  and
\begin{equation} \label{eq:U3}
U_{\alpha} =  P_\alpha G_0^{-1} + P_\alpha  t G_0  U_{\alpha}
\end{equation}
are the  3+1 and 2+2 subsystem transition operators.
$G_0$, $t$, and $U_{\alpha}$ depend on the complex energy parameter $Z$,
but for brevity this dependence is suppressed in the notation.
The bosonic symmetry in the four-body system is imposed by permutation
operators $P_{34}$, $P_1 =  P_{12}\, P_{23} + P_{13}\, P_{23}$, and
$P_2 =  P_{13}\, P_{24} $ where $P_{ab}$ interchanges particles $a$ and $b$.

Symmetrized AGS equations (\ref{eq:U}) are solved in the momentum-space partial-wave
representation $|k_x k_y k_z [(l_x l_y)J l_z] \mathcal{JM} \rangle_\alpha$,
where $k_x, k_y, k_z$ are the magnitudes of the Jacobi momenta \cite{deltuva:12a} and
$l_x, l_y, l_z$ are the associated orbital angular momenta, that are coupled via the intermediate 
subsystem angular momentum $J$ to the total four-body angular momentum
$\mathcal{J}$ with the projection $\mathcal{M}$. In the context of Efimov tetramers
$\mathcal{J}=0$. Furthermore, given the universality, for the numerical efficiency
it is convenient to restrict the two-body interaction to the $s$-wave, i.e., $l_x=0$,
and to take it of a separable form
\begin{equation} \label{eq:lsep}
  \langle k'_x l_x | v | k_x l_x \rangle = \delta_{l_x 0}
  e^{-(k'_x/\Lambda)^2} \, \frac{2}{\pi m}
\left\{ \frac1a - 
\frac{\Lambda}{\sqrt{2\pi}} \right\}^{-1} \!\!\! e^{-(k_x/\Lambda)^2},
\end{equation}
where $m$ is the boson mass and $\Lambda$ the momentum cutoff parameter.
With these constrains $l_y=l_z=J$ and formally is unlimited from above, however,
practical calculations reveal that $l_y,l_z,J \le 1$ is sufficient for an accuracy 
better than 1\%. 
Although Ref.~\cite{deltuva:12a} included higher waves $l_y,l_z,J \le 2$
and thereby achieved accuracy better than 0.1\%, the corresponding extension in the present work
would not significantly improve the extraction of resonance parameters,
since they have larger error bars due to other reasons as shown and  discussed 
in the next section.

In the region of negative two-boson scattering length $a$ there are no bound dimers, and 
the singularities in the kernel of the  AGS equations (\ref{eq:U}) arise due to the Efimov trimer
bound state poles in $U_1$ and due to free resolvent $G_0$. Their treatment using the
complex-energy method with special integration weights is taken over from
Ref.~\cite{deltuva:12c}. After the discretization of momentum variables the system
of integral equations (\ref{eq:U}) becomes a system of linear algebraic equations.
Since the spectrum of the four-boson system near unitarity is more rich compared to the
four-nucleon system, it is advisable to solve the linear system by a direct
matrix inversion as in other four-boson studies \cite{deltuva:12a,deltuva:12d}
based on the integral equations for transition operators. This is possible
taking advantage of a simple potential form (\ref{eq:lsep}) that allows 
to reduce the number of continuous variables in the AGS equations (\ref{eq:U}).
References \cite{deltuva:12a,deltuva:12d} provide more details and also relations of transition
operators $\mcu_{\alpha \beta}(Z)$ to scattering amplitudes and observables.

\section{Results \label{sec:res}}

The present work studies the evolution of four-boson Efimov resonances
in the $a<0$ two-boson scattering length region above the
four-body breakup threshold. For this purpose the scattering length $a$ is
varied in Eq.~(\ref{eq:lsep}); $|a|$ decreases when decreasing the two-boson attraction 
and the system moves away from the unitary limit.
Special value of $a$ where the $n$-th Efimov
trimer crosses the $E=0$ threshold is denoted by $a_n^-$, with $n=0$ labeling
the ground state and $n\ge 1$ the excited states. In the bound state regime there are two Efimov tetramers associated
with each trimer, they are labeled with two integer numbers $n,k$, where $k=1$ (2) corresponds
to a more (less) tightly bound tetramer, called  also deep (shallow) tetramer in the literature.
Their intersections with the $E=0$ threshold are labeled by  $a_{n,k}^-$.
Although in the resonance regime the wording deep or shallow is not really meaningful, it will be employed nevertheless
in order to relate the resonances to the states from which they evolved.
In the unitary limit $a_{n+1}^-/a_n^- \approx 22.694$, the universal Efimov ratio \cite{braaten:rev}.
Furthermore, $a_{n,1}^-/a_n^- = 0.4254(2)$ and  $a_{n,2}^-/a_n^- = 0.9125(2)$
as determined in accurate numerical calculations in Ref.~\cite{deltuva:12a}.
In fact, already for $n=2$ with the force model of  Eq.~(\ref{eq:lsep}) the
 deviations from the above ratios are well below 0.4\% as shown in Ref.~\cite{deltuva:12a}.
For a fixed $n$ the error in $a_{n,k}^-$  due to the limitation $l_y,l_z,J \le 1$ is even smaller, well below 0.1\%.
Together with the convergence study in  Refs.~\cite{deltuva:12d,deltuva:20a} this suggests that
$n=2$ and $l_y,l_z,J \le 1$ is sufficient for the extraction of the universal results with a good accuracy, better than 1\%.
A further justification is provided by the fact that in this regime the two-boson effective range
$r_s$ is already much smaller than $a$, 
the ratio  $|r_s/a|$ that quantifies the finite-range corrections
being of the order of $0.001$ for $n=2$. For comparison, in the regime relevant for
the $n=1$ states $|r_s/a|$ is of the order of $0.02$. 
Alternatively, large values of $|a \Lambda| $ also show that $a$ largely exceeds the interaction range;
$ 1020 <  |a \Lambda| < 3776 $ in the present calculations for $n=2$.

The universal character of the results becomes more evident when represented in dimensionless
quantities. The reference point in $a$ for each Efimov tetramer is chosen as
$a = \ankz$ that connects the unstable bound state and resonance regimes.
Preserving the consistency with the standard representation of the Efimov 
physics in terms of $1/a$, the results will be given as functions of the dimensionless ratio
$|\ankz|/a$, with $|\ankz|/a > -1$ corresponding to the unstable bound state
while $|\ankz|/a < -1$ in the resonance regime.
Furthermore, the energy $E^R_{n,k}$ and width $\Gamma_{n,k}$ of the $(n,k)$-th Efimov tetramer
will be presented in the dimensionless forms
$\varepsilon_{n,k} = E^R_{n,k} \, m (\ankz)^2/\hbar^2 $
and $\gamma_{n,k} =  \Gamma_{n,k} \, m (\ankz)^2/\hbar^2$.
Note that, up to the finite-range corrections, $-\hbar^2/m (\ankz)^2$ is the
energy of the virtual two-boson state at the crossing point $a = \ankz$.
Thus, $\varepsilon_{n,k}$ ($\gamma_{n,k} $) is the energy (width)
of the  $(n,k)$-th Efimov tetramer in units of the dimer virtual state energy taken at
the reference point $a = \ankz$.

\begin{figure}[!]
\begin{center}
\includegraphics[scale=0.64]{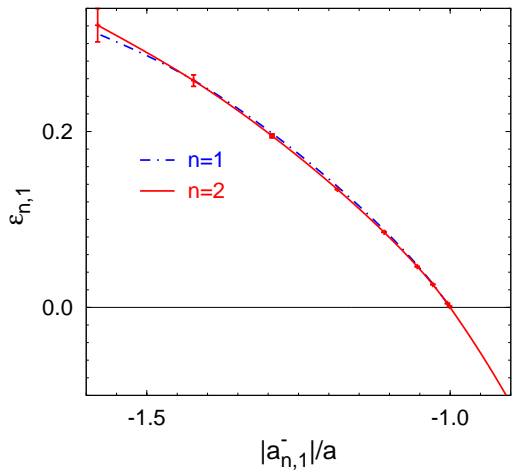}
\end{center}
\caption{\label{fig:Edeep} (Color online)
Dimensionless real energy part of the deep ($k=1$) Efimov tetramer 
as a function of the inverse two-boson scattering length.
The results for the states associated with 
the first and second excited Efimov trimer are displayed by
dashed-dotted and solid curves, respectively.
}
\end{figure}

\begin{figure}[!]
\begin{center}
\includegraphics[scale=0.64]{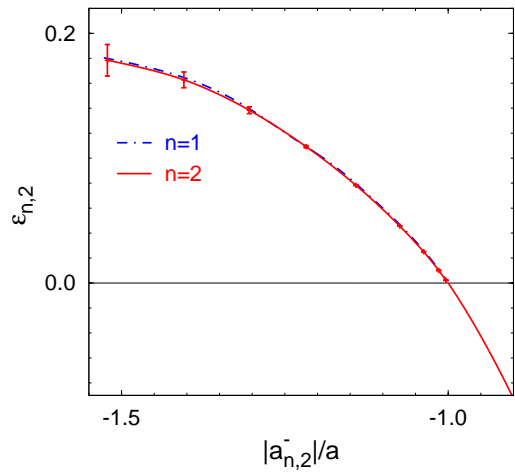}
\end{center}
\caption{\label{fig:Eshl} (Color online)
Dimensionless real energy part of the shallow ($k=2$) Efimov tetramer 
as a function of the inverse two-boson scattering length.
Curves are as in Fig.~\ref{fig:Edeep}.
}
\end{figure}

\begin{figure}[!]
\begin{center}
\includegraphics[scale=0.64]{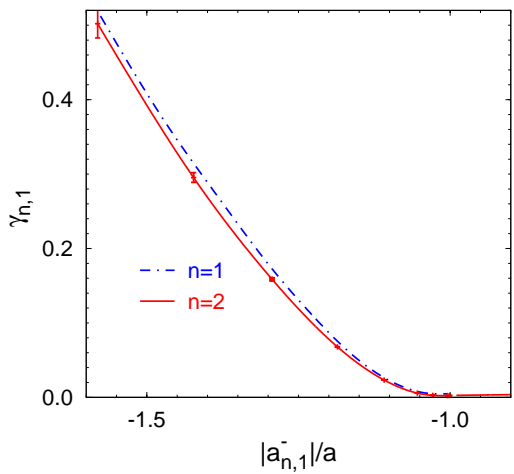}
\end{center}
\caption{\label{fig:Gdeep} (Color online)
Dimensionless width of the deep ($k=1$) Efimov tetramer 
as a function of the inverse two-boson scattering length.
Curves are as in Fig.~\ref{fig:Edeep}.
}
\end{figure}

\begin{figure}[!]
\begin{center}
\includegraphics[scale=0.64]{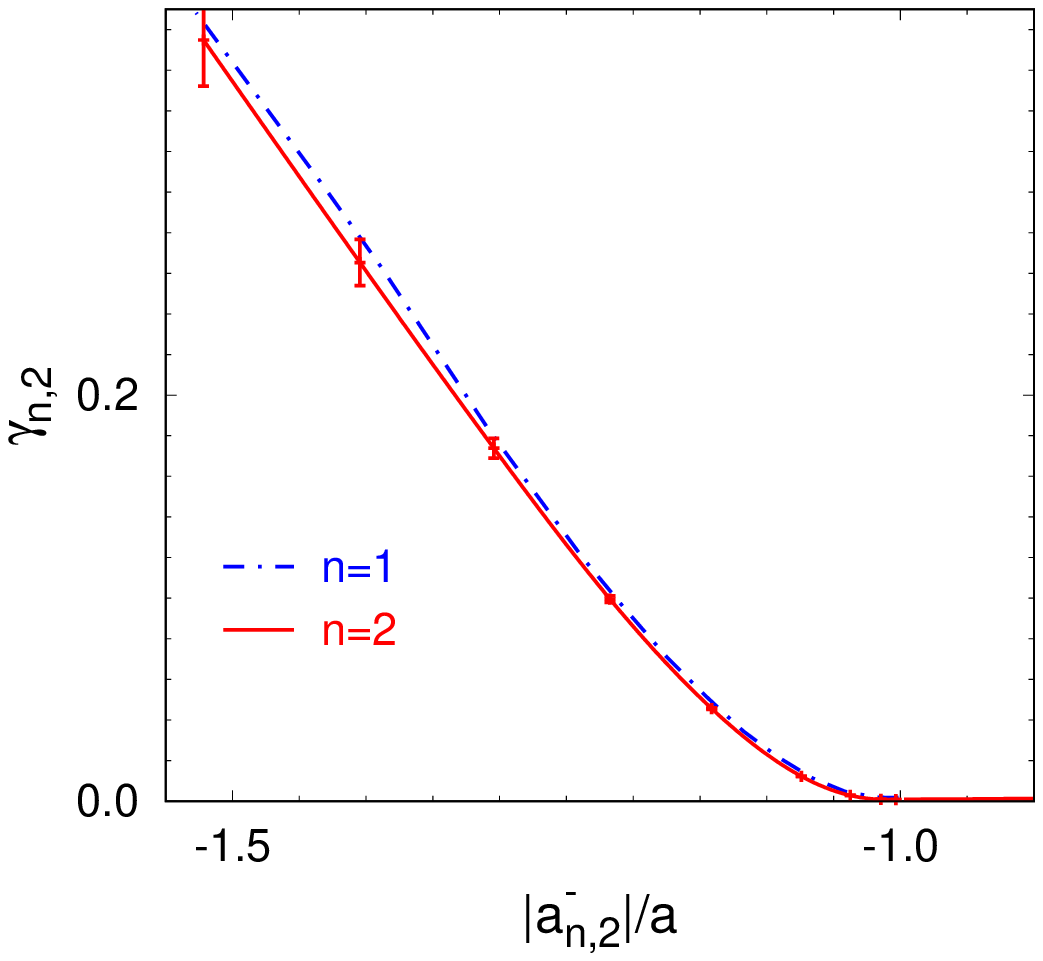}
\end{center}
\caption{\label{fig:Gshl} (Color online)
Dimensionless width of the shallow ($k=2$) Efimov tetramer 
as a function of the inverse two-boson scattering length.
Curves are as in Fig.~\ref{fig:Edeep}.
}
\end{figure}

The dimensionless energies $\varepsilon_{n,k}$ of the tetramers associated with
the first two excited Efimov trimers $n=1$ and 2 are shown in  Figs.~\ref{fig:Edeep}
and  \ref{fig:Eshl} for the deep and shallow levels, respectively. For $n=2$ also
the unstable bound state regime is partially given.
In this representation both types of tetramers  
exhibit a qualitatively similar dependence on the scattering length:
when moving away from the unitary limit, i.e., weakening the attraction between bosons,
the unstable bound state turns into a
 resonance whose energy increases but with some saturation.
 The finite-range effects appear to be small also for the $n=1$ level, providing
 confidence on the convergence towards universal limit with increasing $n$.
 The evolution of the corresponding  dimensionless widths $\gamma_{n,k}$ is
presented in  Figs.~\ref{fig:Gdeep}
and  \ref{fig:Gshl} for the deep and shallow tetramer, respectively.
Note that also in the unstable bound state regime the tetramer width is finite,
though small in the absolute size; see Ref.~\cite{deltuva:12d} for a detailed study.
As a consequence, at the transition point  $a = \ankz$ the four-boson resonances
have finite width, in contrast to the zero width in typical cases where the
true bound state evolves into the resonance as, for example, the Efimov trimer
does \cite{deltuva:20a}.
Moving deeper into the resonance region towards smaller $|\ankz|/a$ values
the widths of the four-boson Efimov resonances increase.
Again, the evolution is qualitatively similar for deep and shallow tetramer.
For both of them the differences between $n=1$ and 2 results are
slightly larger than in the $\varepsilon_{n,k}$ case; this is consistent
with the unstable bound state regime where the width shows somehow slower convergence
with $n$ \cite{deltuva:12d}.
 Larger finite-range effects for the width as compared to the energy
 are not surprising since for the width decisive are the transitions to states of a
particle plus lower-lying
trimer  that obviously are more affected by range corrections.

The symbols in Figs.~\ref{fig:Edeep} - \ref{fig:Gshl} show  $n=2$ results
with theoretical error bars (for $n=1$ the error bars are of the same size),
estimated in the same way as in previous works
\cite{deltuva:18b,deltuva:20a}. For narrow and well pronounced resonances
close to  $|\ankz|/a = -1$ the
errors are very small. However, as the  width of  the resonance increases, the
nonresonant background terms in the transition operators become dominant,
the resonant behavior can hardly be seen, which results in large uncertainties
in the determination of resonance parameters. For this reason the present
results are limited to $|\ankz|/a > -1.6$, however, beyond this limit the
resonant behavior in the four-boson continuum is disappearing
and becomes physically unobservable.

\begin{figure}[!]
\begin{center}
\includegraphics[scale=0.68]{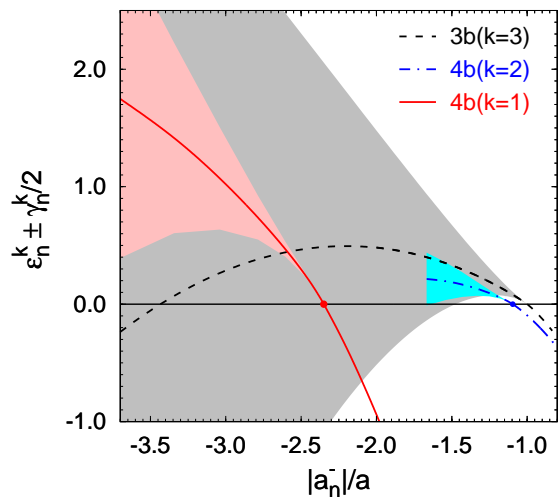}
\end{center}
\caption{\label{fig:e34} (Color online)
Energies of three- and four-boson Efimov resonances
as functions of the inverse two-boson scattering length.
The curves represent the $\varepsilon_{n}^{k}$ values 
while the shaded areas around each curve extend from
$\varepsilon_{n}^{k} - \gamma_{n}^{k}/2$ to
$\varepsilon_{n}^{k} + \gamma_{n}^{k}/2$ and thereby
reflect the width of the respective resonances.
The two points at the threshold correspond to 
$a = \ankz$ with $k=1$ (left) and 2 (right), respectively.
}
\end{figure}

Given the proximity of $\anz$ and $a_{n,2}^-$, the three-boson Efimov resonances
appear in a nearby regime. It is interesting to compare three- and
four-boson Efimov resonances using the same scale. For this purpose
$\varepsilon_{n}^{k} = E^R_{n,k} \, m (\anz)^2/\hbar^2 $
and $\gamma_{n}^{k} =  \Gamma_{n,k} \, m (\anz)^2/\hbar^2$ are introduced, 
where $k=3$ formally corresponds to the trimer; in this case the 
results are taken from Ref.~\cite{deltuva:20a}.
In other words, $\varepsilon_{n}^{k}$ and  $\gamma_{n}^{k}$
are  energies and  widths of the respective few-boson Efimov states
in units of the dimer virtual state energy taken at 
the common reference point $a = \anz$ where the trimer becomes unbound.
The results for the trimer and tetramer energies are displayed in Fig.~\ref{fig:e34}
by solid, dashed-dotted, and dashed curves for $k=1$, 2, and 3, respectively.
To reflect the width of each state, the shaded areas around each curve
cover the values between  $\varepsilon_{n}^{k} - \gamma_{n}^{k}/2$
and  $\varepsilon_{n}^{k} + \gamma_{n}^{k}/2$.
The three-boson resonance broadens most
rapidly and evolves into a physically unobservable  subthreshold resonance 
with $\varepsilon_{n}^{3} < 0$ residing in the third quadrant of the complex
energy plane. The trajectory of the shallow tetramer is roughly parallel to the trimer,
just shifted to more negative $|\anz|/a$ values and with lower energy and width.
The complexity of four-body calculations precludes a reliable parameter extraction 
of very broad and practically unobservable resonances, but the similarity to the trimer
case suggests the conjecture that the $k=2$ tetramer evolves into the 
subthreshold resonance as well when moving away from the unitary limit.
Note that this is a typical behavior of resonances observed in 
a number of other few-body systems
\cite{PhysRevC.66.054001,lazauskas:3n,bringas:04a,lazauskas:4n,PhysRevC.89.032201}.
The deep tetramer evolves into a resonance with the highest energy, exceeding also the trimer.
However,  one has to keep in mind that there is no real crossing of these states since, given very different widths, 
in the complex energy plane they are far from each other; the  trimer is very broad and does not
affects the physical four-boson processes near the $a \approx a_{n,1}^-$ regime.

\begin{figure}[!]
\begin{center}
\includegraphics[scale=0.68]{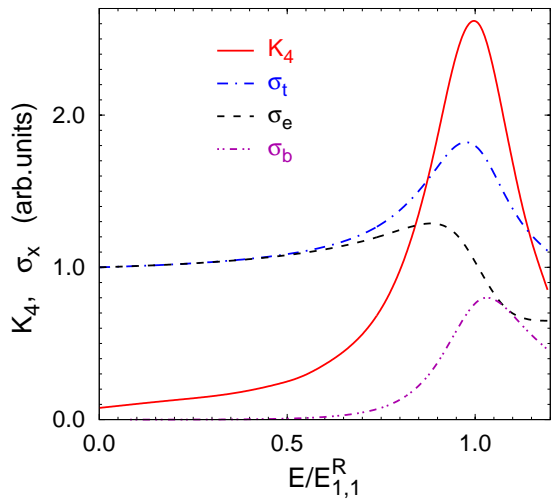}
\end{center}
\caption{\label{fig:obs} (Color online)
Four-boson recombination rate $K_4$ and cross section components
for the boson-trimer scattering as  functions of the system energy.
The resonance corresponds to $(n,k)=(1,1)$ with $|a_{1,1}^-|/a = -1.1035$, 
$\varepsilon_{1,1} = 0.433$, and $\gamma_{1,1}/\varepsilon_{1,1} = 0.286$.
}
\end{figure}

The present method for determining the resonance parameters from the energy dependence of physical
transition operators (scattering amplitudes) is not applicable to physically unobservable  very broad
or subthreshold resonances, in contrast to some methods based on the  complex scaling or analytic continuation in the
coupling constant \cite{lazauskas:3n,lazauskas:4n}.
However, it has an advantage of predicting simultaneously also the scattering observables, including both
resonant contributions and nonresonant background, whose relative importance determines to what extent
the resonant behavior is pronounced and can be observed. Figure~\ref{fig:obs} presents an example
for $(n,k)=(1,1)$ at $a = 0.4\, a_{1}^- = 0.9062 \, a_{1,1}^-$. Several collision channels are possible:
(i) four free bosons can recombine into the ground-state Efimov trimer plus boson;
(ii) a boson may scatter  from the ground-state Efimov trimer elastically or
(iii) inelastically, i.e., leading to the trimer breakup into three free bosons.
Note that (i) is the time reverse of (iii). The characteristic observables, i.e.,
the four-particle recombination rate $K_4$ \cite{deltuva:12a} and 
the elastic $\sigma_e$, breakup $\sigma_b$, and total $\sigma_t = \sigma_e + \sigma_b$
 boson-trimer scattering cross sections  are shown as functions of the energy. 
Noteworthy, although the resonance energy and width extracted from different amplitudes
agree very well, the relative importance of  resonant and background terms is very 
different for different reactions as Fig.~\ref{fig:obs} demonstrates;
the resonance is most clearly pronounced in the 
four-boson recombination rate $K_4$. 

\section{Summary \label{sec:sum}}

The four-boson continuum was studied  at negative values of the two-boson scattering length.
Reducing its magnitude and
thereby moving away from the unitary limit, the Efimov tetramers evolve from unstable bound states  into resonances.
Exact scattering equations for the four-particle transition operators were solved using the momentum-space techniques.
The energy dependence of various scattering amplitudes was explored  to determine resonance positions and widths.

Using a simple rank-one separable potential, the tetramers associated with the first two excited
Efimov trimers were considered. It was demonstrated that in a proper dimensionless representation
the results for the two levels are close. Taking into account the convergence rate for a number of other
four-boson quantities it was argued that $n=2$ level results approximate well the universal limit,
the remaining finite-range corrections being well below 1\%.

In contrast to the Efimov trimers, even at the transition points
$a = \ankz $ the tetramers (except for $n=0$) have finite  width. However,
the width reaches much higher values when going deeper into the resonance regime.
Nevertheless, the tetramers broaden less rapidly than the trimer.
The observed resonance evolution, when reducing the two-boson attraction and therefore $|a|$, is
typical also in many other systems: the real energy part rises with the decreasing rate while the
width rises with the increasing rate, until the resonance becomes very broad,
physically unobservable, and cannot be followed anymore by the present method.
The seen similarity between the trimer and shallow tetramer evolution suggests
that also the latter evolves into a physically unobservable subthreshold resonance.
In the physically observable regime the shallow tetramer is lower in energy than the trimer
while the deep tetramer can rise to higher energy values.

The advantage of the transition operator method is the ability to predict the physical
observables. Presented examples
demonstrate that the relative importance of resonant and background contributions
depends strongly on the considered observable, and suggest that the  four-particle
recombination rate  exhibits the most pronounced resonant behavior.

\vspace{1mm}

The author acknowledges  support  by the Alexander von Humboldt Foundation
under grant no. LTU-1185721-HFST-E.


\end{document}